\documentclass[nofootinbib,secnumarabic]{revtex4}

\usepackage{graphicx,epsf,amssymb,amsmath,latexsym,amsfonts,ifthen,calc, slashed,feynmf}
\usepackage{subfigure}
\usepackage{epsfig}
\usepackage{lastpage}
\usepackage{fancyhdr}
\usepackage{float}
\usepackage{titletoc}

\usepackage{indentfirst}

\begin{document}

\title{The influence of the Al stabilizer layer thickness on the normal zone propagation velocity in high current superconductors}


\author{I. Shilon}
\affiliation{European Organization for Nuclear Research (CERN), CH-1211, Gen\`eve 23, Switzerland}

\author{A. Dudarev}
\affiliation{European Organization for Nuclear Research (CERN), CH-1211, Gen\`eve 23, Switzerland}

\author{S. A. E. Langeslag}
\affiliation{European Organization for Nuclear Research (CERN), CH-1211, Gen\`eve 23, Switzerland}

\author{L. P. Martins}
\affiliation{European Organization for Nuclear Research (CERN), CH-1211, Gen\`eve 23, Switzerland}

\author{H. H. J. ten Kate}
\affiliation{European Organization for Nuclear Research (CERN), CH-1211, Gen\`eve 23, Switzerland}

\begin{abstract}
The stability of high-current superconductors is challenging in the design of superconducting magnets. When the stability requirements are fulfilled, the protection against a quench must still be considered. A main factor in the design of quench protection systems is the resistance growth rate in the magnet following a quench. The usual method for determining the resistance growth in impregnated coils is to calculate the longitudinal velocity with which the normal zone propagates in the conductor along the coil windings.

Here, we present a two dimensional numerical model for predicting the normal zone propagation velocity in Aluminum stabilized Rutherford NbTi cables with large cross section. Such conductors comprise a superconducting cable surrounded by a relatively thick normal metal cladding. By solving two coupled differential equations under adiabatic conditions, the model takes into account the thermal diffusion and the current redistribution process following a quench. Both the temperature and magnetic field dependencies of the superconductor and the metal cladding materials properties are included. Unlike common normal zone propagation analyses, we study the influence of the thickness of the cladding on the propagation velocity for varying operating current and magnetic field. 

To assist in the comprehension of the numerical results, we also introduce an analytical formula for the longitudinal normal zone propagation. The analysis distinguishes between low-current and high-current regimes of normal zone propagation, depending on the ratio between the characteristic times of thermal and magnetic diffusion. We show that above a certain thickness, the cladding acts as a heat sink with a limited contribution to the acceleration of the propagation velocity with respect to the cladding geometry. Both numerical and analytical results show good agreement with experimental data.
\end{abstract}

\maketitle


\section{Introduction}

In superconducting magnets, the design of an adequate quench protection system based on an internal dump of the magnet stored energy depends mainly on the growth rate of the resistance that builds up in the magnet following a quench. To determine the resistance growth rate, one usually turns to study the initiation and dynamics of the normal zone, the region within which the superconductor exhibits normal behavior, in the conductor. The normal zone propagation in adiabatic conditions can be described by two parameters, the longitudinal velocity $v_l$, the rate with which the normal zone expands along the axis of the conductor, and the transverse velocity $v_t$. $v_l$ is much easier to measure in an experiment and a computation of it can be done either numerically or by utilizing some of the analytical formulae available in the literature (see, for example, \cite{ansysquench,hermanquench,devred1}). Moreover, $v_t$ can be linearly approximated by $v_t = v_l\sqrt{k_t/k_l}~,$ where $k_t$ and $k_l$ are the transverse and longitudinal thermal conductivities of the superconductor, respectively \cite{willi}. Hence, a satisfying description of the normal zone propagation can be obtained  by knowing $v_l$.

In this work, we present a novel numerical calculation of $v_l$ in NbTi/Cu Rutherford cables surrounded by a normal metal cladding. Such conductors are used in many existing detector magnets, such as the famous ATLAS and CMS magnets at CERN, and will be utilized by the future IAXO experiment \cite{iaxojinst}. The numerical calculation exploit the commercial FEA software COMSOL to simulate the propagation of a normal zone in a two dimensional adiabatic conductor. The model accounts for both the current sharing process between the superconductor and the stabilizer, as well as for the heat propagation over time and space along the conductor. Hence, it yields the influence of both the temperature $T$ and the magnetic field $B$ on $v_l$. In addition, we study the influence of the thickness of the cladding on $v_l$ for varying magnetic field and operating current. This allows us to present a good estimation of the longitudinal normal zone propagation velocity for a very broad variety of highly stabilized superconductors in many existing and also, more importantly, for future magnets.  

To complete our analysis, we introduce an analytical formula to calculate $v_l$, following a previous idea by Mints et al. \cite{mints}. This formula allows one to approximate $v_l$ by taking into account the thermal diffusion as well as the current redistribution in the conductor. We apply the formula in the same scenarios as in the numerical model to aid us in analyzing and interpreting the results of the numerical calculation and also to give a further justification to the numerical results when experimental data is unavailable. 

\section{Computational Model}
\label{compmod}

In the core of the numerical computation is a solution to the heat diffusion equation that takes into account the current redistribution process in the conductor. The heat balance equation for a unit volume of conductor is given by 

\begin{equation}
\label{heatbalance}
\displaystyle \varrho C(T)\frac{\partial T}{\partial t}-\vec{\nabla}\cdot (k(T)\vec{\nabla}T )= \rho(T)|\vec{J}(t)|^2 + q_{ext}(\vec{r}, t)~,
\end{equation}

\noindent where $\varrho$ is the mass density, $C$ is the specific heat, $k$ is the thermal conductivity, $\rho$ is the electrical resistivity, $J$ is the current density and $q_{ext}(\vec{r}, t)$ represents the external energy disturbance. As the temperature in the conductor rises above the current-sharing temperature, the current starts diffusing from the superconductor into the normal metal, thereby continuously changing the heat generation term in Eq. (\ref{heatbalance}). To include this effect in the heat generation term, one simply introduces Ampere's law into the heat balance equation. 

Assuming the electromagnetic behavior of the conductor is similar to that of a set of infinite plates (see Sec. \ref{modgeo}), we regard the current carrying cable as an infinite current carrying sheet. Thus, the displacement current is omitted from the Maxwell-Ampere equation and one can easily obtain an equation to describe the magnetic diffusion 

\begin{equation}
\label{difM}
\displaystyle\vec{\nabla}\times\left(\rho(\vec{r}, t)\vec{\nabla}\times\vec{B}(\vec{r}, t)\right)+ \mu_0\frac{\partial \vec{B}(\vec{r}, t)}{\partial t} = 0~.
\end{equation}

\noindent From the coupled set of Eqs. \ref{heatbalance} (utilizing Ampere's law) and \ref{difM}, a full description of the temperature and magnetic field distributions in a current carrying object can be obtained.

As we seek for the behavior of $v_l$ as a function of $T$ and $B$, we must account for the dependence of the material properties of both these parameters. To relax the computational cost, we assume the Rutherford cable is made of a single, homogenous, material and average $\varrho,~C$ and $k$ over the cross-section of the cable. The effective electrical density is obtained by viewing the superconductor-copper system as two resistors connected in parallel. The material properties of the cladding are those of the normal metal comprising it. The effective resistivity of the conductor is obtained by treating the cladding and Rutherford cable as two resistors connected in parallel. The material properties were obtained from the MATPRO library~\cite{matpro} and fitted to a polynomial function. The different fits cover a magnetic field range of 0-5~T for a temperature range of 0-300~K. The Residual Resistivity Ratio (RRR) we chose for the aluminum in the cladding and the copper are 1500 and 100, respectively.

\subsection{Model Geometry and Mesh}
\label{modgeo}

The minimal dimensionality of quench study models depends on the ratio between the time scales of the thermal and magnetic diffusions. The thermal diffusion time scale is given by $\tau_H=L_{nzf}^2/D_H$ , where $L_{nzf} = v\tau_H$ is the length of the region in which the quench-driving heat release occurs, $v$ is the propagation velocity and $D_H = k/\varrho C$ is the thermal diffusivity of the cladding, taken at $T_{cs}$.  The time scale of the transverse magnetic diffusion is defined as $\tau_M=\frac{\mu_0 L_{2}^2}{\rho_{st}}$ , where $L_2$ is the thickness of the cladding and the index $st$ refers to cladding metal. As $\tau_H$ depends implicitly on the current through the velocity, for high currents $\tau_M$ is generally finite with respect to $\tau_H$. Thus, a sufficient understanding of the problem requires a 2D model. Notice, however, that a solution to this problem can also be obtained by coupling a 2D magnetic model to a 1D thermal model.

\begin{figure}
\centering
\subfigure[~Cross section of the conductor model.]{\includegraphics[width=6cm]{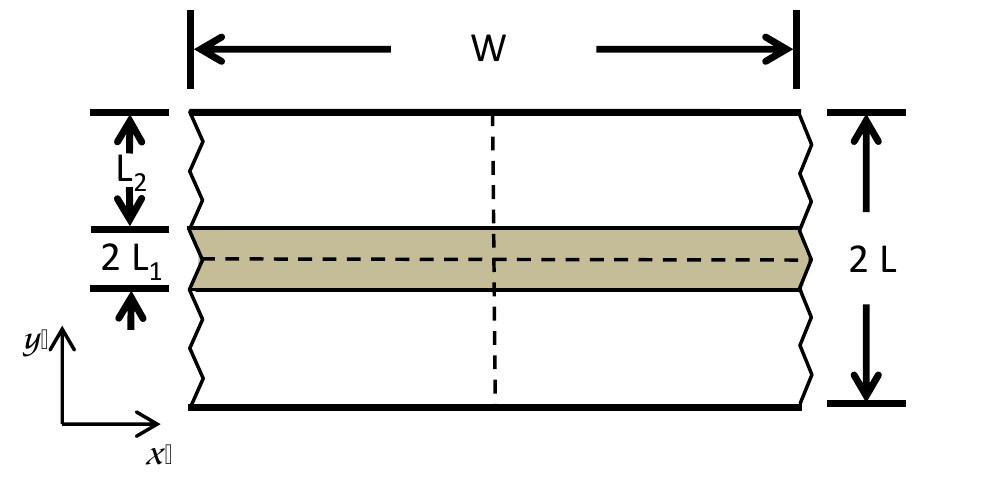}\label{BTgeom}}
\subfigure[~The 2D geometry used for the numerical analysis.]{\includegraphics[width=10cm]{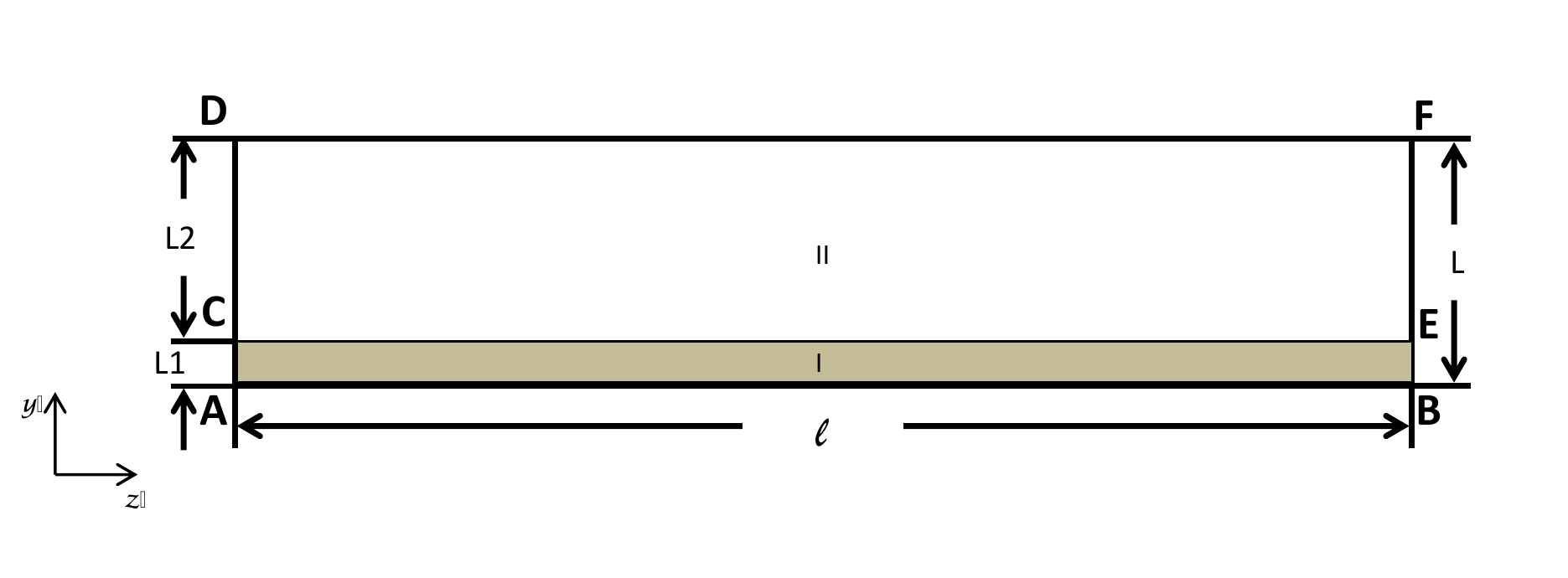}}
\caption{The conductor geometry used to model the normal zone propagation in a 2D model. (a) The cross-section of the conductor. (b) The 2D model is solved over a confined region in the $yz$ plane.}
\label{geometria}
\end{figure}

The 2D model of the conductor is shown in Fig. \ref{geometria}. We assume that the length scale in the $\hat{x}$ direction is infinitely larger than any other length scale in the problem and the reduction to an equivalent 2D problem is thus done by considering a slice of the conductor on the $yz$ plane. The solution is restricted to the $0\leq y \leq L$ and $0\leq z \leq l$ region. In the $y$ direction, we refer to the thickness of the cable as $L_1$ and to that of the cladding as $L_2$.

\begin{figure}[!h]
\includegraphics[width=11cm]{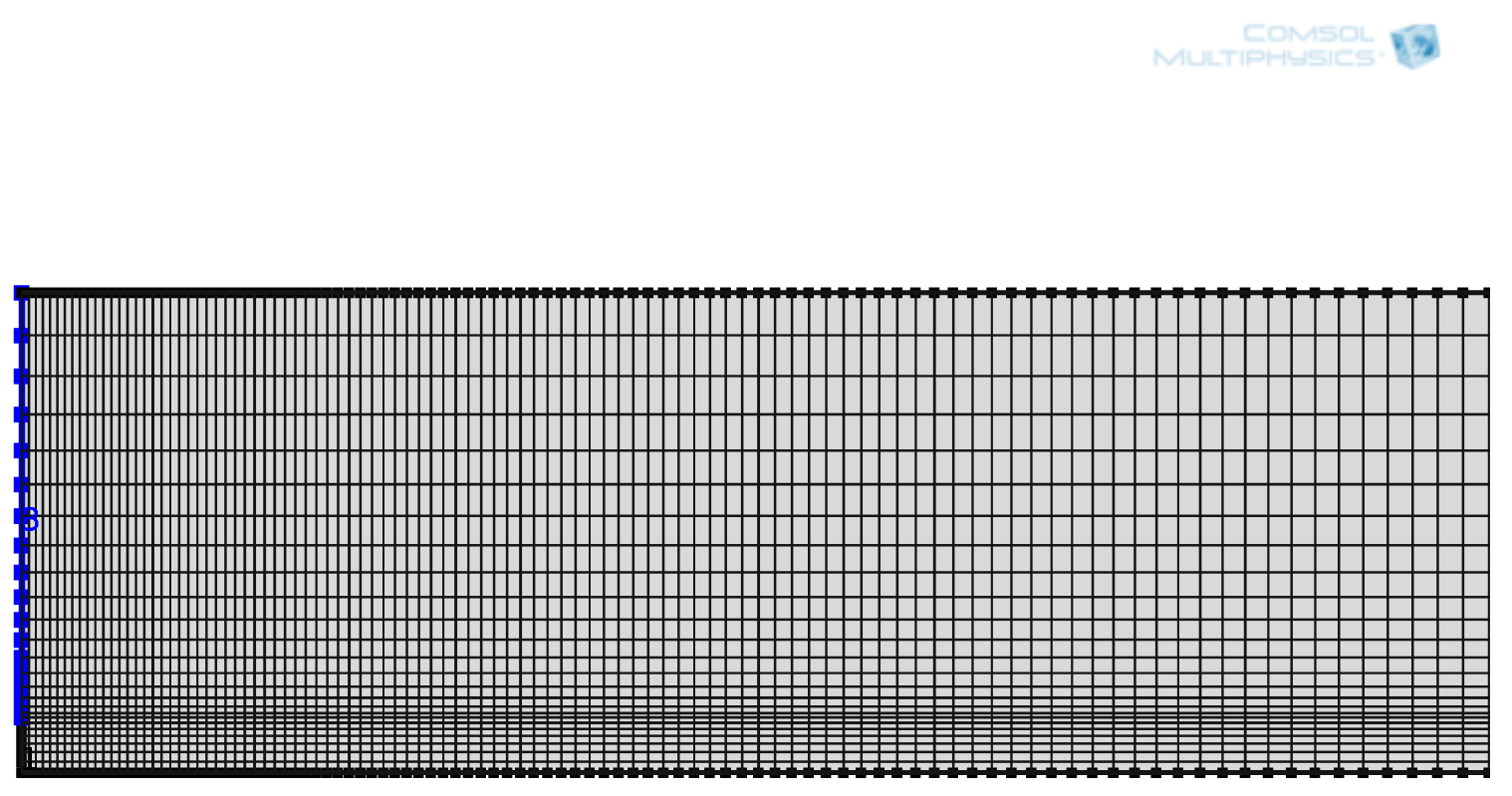}
\caption{An illustration of the mesh used for the COMSOL simulation.}
\label{comsolmesh}
\end{figure}

The mesh of the geometry uses rectangular 2D elements and is shown in Fig. \ref{comsolmesh}. The elements are constructed so that they are finer in the vicinity of the boundary between the cable and the cladding, at $L_1$. In the $z$ direction the elements are getting coarser with increasing $z$, so that near the origin, where the initial perturbation takes place and the temperature gradient is high, the mesh is considerably finer. The model consists of 11000 domain elements and 2022 boundary elements for $l = 2$~m.

\subsection{Boundary Conditions and Initial Perturbation}

An immediate consequence of the 2D description of the problem is that the conductor can be approximated by a set of infinite plates, where the cable is seen as an infinite current carrying sheet with initial total current $I_{op}$. Thus, the magnetic field in the conductor prior to the external energy release has the form

\begin{equation}\label{B0}
B(t=0) = B_0 = B_{\text{ext}}+\left\{\begin{array}{cc}
\mu_0 \frac{I_{\text{op}}}{A_{\text{R}}}y&y<L_1~,\\
\mu_0 J_{\text{eng}} L & y\ge L_1~,\\
\end{array}\right.
\end{equation}

\noindent where $J_{\text{eng}} = I_{\text{op}}/A_{\text{total}}$ is the engineer current density an $B_{ext}$ is an external magnetic field representing the contribution from other current sources in the coil. On the interface $\Gamma$ between the composite material and the cladding we demand the continuity of the magnetic field and its flux. On the external boundaries, the magnetic field is defined as $B(y=0) = B_{\text{ext}}$ and $B(y=L) = B_{\text{ext}} + \mu_0 J_{\text{eng}} L$. 

The model assumes full adiabaticity. The initial value condition for the temperature reads $T(\vec{r}, t=0) = T_{op}$ in the bulk of the conductor. The temperature and its flux are also assumed to be continuous along $\Gamma$, so that $\hat{n}\cdot( k_I \vec{\nabla} T_I-k_{II} \vec{\nabla} T_{II})=0$~, and that $T_I = T_{II}$, where the indices $I$ and $II$ correspond to the composite material region and the cladding region, respectively, and $\hat{n}$ is the unit vector orthogonal to the boundary.

The external energy input can be computes in several ways. We chose to represent the energy pulse by a Guassian shape in space and exponential decay in time. The disturbance is given by a power density function

\begin{equation}
P_{\text{ext}} = P_0 ~ \mbox{e}^{-(z^2 + y^2)\cdot\sigma}\mbox{e}^{-t^2/\tau^2}~,  
\end{equation}

\noindent where $\sigma = 4 \log 2 /d^2$ with $d = 10$ mm being the full width at half maximum of the Gaussian, and $\tau$ = 0.005 sec. From $E_Q = \int_{-\infty}^\infty dz dy \int_0^\infty dt P_{\text{ext}}$, where $E_Q$ is a 1D energy density, we get $P_0 = 2 E_Q \sigma / \pi^{3/2} \tau$ W/m$^3$. Thus, the initial quench energy can be estimated by multiplying $E_Q$ by an appropriate characteristic length, such as the coil width.


\section{Analytical Approximation}
\label{analyticalmodels}

Next, it is worth writing an analytical formula to describe the longitudinal normal zone propagation. Although due to the strong coupling between the heat and magnetic diffusion equations an exact solution is practically impossible, a good approximation can be obtained in a simple manner. 

A well known technique to deal with the heat diffusion equation in a 1D adiabatic system can be found in \cite{willi}. Although current redistribution is not taken into account, this technique can provide a good first approximation for low currents, where the current redistribution can be regarded as instantaneous. In this case, a solution to the heat diffusion equation yields the following longitudinal propagation velocity $v_{\text{wil}}=\frac{I_{op}}{\varrho CA_{\text{cd}}}\sqrt{\frac{\rho k}{T_{cs} - T_{op}}}$~, where $A_{cd}$ is the total cross-section of the conductor and the material properties are taken as an average across $A_{cd}$.

When the operating current is high, the assumption that the current is immediately redistributed into the cladding is no longer valid. To distinguish between the high and low current regimes we look at the ratio between the characteristic times associated with the normal zone propagation and the magnetic diffusion \cite{mints}

\begin{equation}
\label{ }
	\alpha = \frac{\tau_{H}}{\tau_{M}} = \frac{k \rho_{st}}{\varrho C v^2 \mu_0 L_{2}^2}~.
\end{equation}


We define the low current regime, where the current can be regarded as immediately redistributed into the cladding, for $\alpha > 1$. Similarly, when $\alpha < 1$ current redistribution must be explicitly taken into account. Then, in the vicinity of the transition front the current remains confined to a certain small fraction of the cladding around the superconductor. This leads to a non-uniform quench-driving heat release, which accelerates the propagation velocity. The transition $\alpha = 1$ takes place at around 10~kA for most highly stabilized cables at their operating points.


A simple way to approximate this scenario is by considering the joule heating term as resulting from a uniform current flowing solely in a confined area $A_{eff}$ within the conductor \cite{mints}. This ansatz introduces the effect of current redistribution into the propagation velocity by solving only the heat balance equation. To find an expression for $A_{eff}$ one may study its asymptotic behavior. The cross-section area of the region where the current flows is determined by $\alpha$. When $\alpha$ is large the current is practically instantaneously redistributed in the conductor and we expect that $A_{eff} \rightarrow A_{cd}~,~\text{for}~ \alpha \rightarrow \infty$~. On the other hand, for small $\alpha$ the current penetrates only a thin layer of the cladding around the cable. Therefore, $A_{eff} \rightarrow A_{R}~,~\text{for}~ \alpha \rightarrow 0$~.

In a similar manner to Boxman et al. \cite{boxman}, we suggest the following expression for the effective area $A_{eff}$ which carries the current

\begin{equation}
\label{aeff}
	A_{eff}=2W\left[L_2\left(1-e^{-\alpha}\right)+L_1\right]~.
\end{equation}

When considering the joule heating to be generated only within $A_{eff}$, the expression that describes the normal zone propagation velocity changes accordingly and takes the following form 

\begin{equation}\label{vMB}
v_{MB}=\frac{I_{op}}{\varrho C\sqrt{A_{\text{eff}} A_{\text{cd}}}}\sqrt{\frac{k \rho_{\text{eff}}}{T_{cs}-T_{op}}}~,
\end{equation}

\noindent where $\rho_{e_{eff}}$ is the effective electrical resistivity calculated from $A_{eff}$

\begin{equation}
\label{ }
	\rho_{eff} = A_{eff}\left(\frac{A_{sc}}{\rho_{NbTi}} + \frac{A_{Cu}}{\rho_{Cu}} + \frac{A_{eff} - A_{R}}{\rho_{st}}\right)^{-1}~.
\end{equation}

The material properties in the latter are taken at $T_{cs}$.

\section{Results}
\label{resuts}

The results of the analytical approximation and the numerical COMSOL model are presented along with two sets of measurements data \cite{boxman, b00}. The measurements were done on two highly stabilized conductors, the B0 and B00 coils of the ATLAS Magnet Test Facility at CERN, for different operating currents. This data, however, is not sufficient to fully verify our numerical and analytical models, as we are also interested in the behavior of the velocity for different cladding thicknesses. Nonetheless, The comparison between the models and the data gives a good indication that the model provides satisfying results.

\subsection{Velocity versus Operating Current}

\begin{figure}
 \begin{center}
 \subfigure[~Velocity vs. operating current for the B0 coil case.]{\includegraphics[scale=2.8]{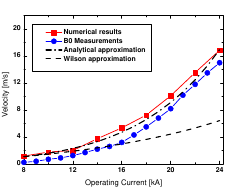}}
 \subfigure[~Velocity vs. operating current for the B00 coil case.]{\includegraphics[scale=2.8]{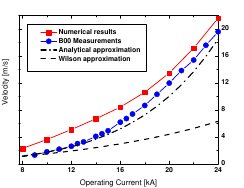}}
 \end{center}
 \caption{A comparison between the Wilson model, the Boxman-Mints model, the COMSOL$^{\circledR}$ simulation and the measurements data from the B0 (left) and B00 (right) coils.} 
 \label{data}
\end{figure}

A measurement of the propagation velocity for different operating currents is a straightforward way to gain insight on the normal zone behavior of the conductor. In Fig. \ref{data}, a comparison between the COMSOL simulation, the measurements data, the analytical formula and Wilson's approximation are presented. The velocity increases exponentially with the current and shows an asymptotic behavior as $I \rightarrow I_c$. The close agreement between the analytical and numerical models and the measurements of the propagation velocity can be appreciated from the graphs.

Another fact evident from the graphs is the breakdown of Wilson's solution for high currents, where current redistribution is becoming significant. Our analytical approximation provides a better match to the data for a wide range of operating currents. One can notice how both Wilson's approximation and our analytical formula converge at low currents, where $A_{eff} \rightarrow A_{cd}$. The simulation results are generally higher than the measurement data due to the adiabatic boundary conditions of the numerical model, which assume the conductor to be a closed system.

\subsection{Velocity versus Aluminum cladding layer thickness}

\begin{figure*}[!t]
\centering
\subfigure[$I_{\text{op}}=10$ kA and $B_{\text{ext}}=4.5$ T]{\includegraphics[width=5.4cm]{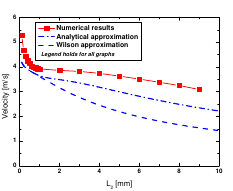}\label{G10kA45T}}
\subfigure[$I_{\text{op}}=10$ kA and $B_{\text{ext}}=1.6$ T]{\includegraphics[width=5.4cm]{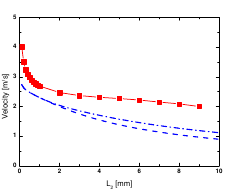}\label{G10kABT}}
\subfigure[$I_{\text{op}}=15$ kA and $B_{\text{ext}}=4.5$ T]{\includegraphics[width=5.4cm]{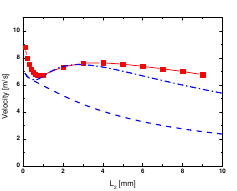}\label{G15kA45T}}
\subfigure[$I_{\text{op}}=15$ kA and $B_{\text{ext}}=2.7$ T]{\includegraphics[width=5.4cm]{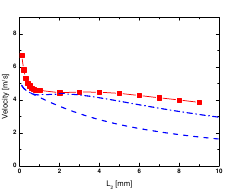}\label{G15kABT}}
\subfigure[$I_{\text{op}}=20$ kA and $B_{\text{ext}}=4.5$ T]{\includegraphics[width=5.4cm]{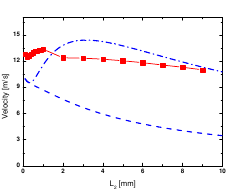}\label{G20kA45T}}
\subfigure[$I_{\text{op}}=20$ kA and $B_{\text{ext}}=3.2$ T]{\includegraphics[width=5.4cm]{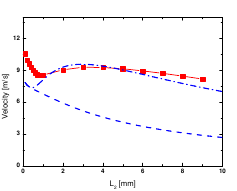}\label{G20kABT}}
\subfigure[$I_{\text{op}}=25$ kA and $B_{\text{ext}}=4.5$ T]{\includegraphics[width=5.4cm]{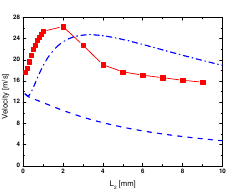}\label{G25kA45T}}
\subfigure[$I_{\text{op}}=25$ kA and $B_{\text{ext}}=4.0$ T]{\includegraphics[width=5.4cm]{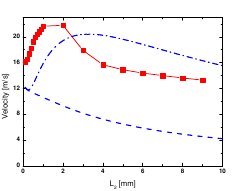}\label{G25kABT}}
\subfigure[$I_{\text{op}}=30$ kA and $B_{\text{ext}}=4.5$ T]{\includegraphics[width=5.4cm]{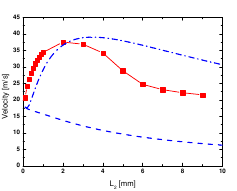}\label{G30kA45T}}
\caption[Velocity as function of Stabilizer thickness.]{Velocity as function of stabilizer thickness for different operating currents and external magnetic fields. The red squares are the results obtained from COMSOL model and the dash dash dot lines are from Mints-Boxman's approximation~\eqref{vMB}.}
\label{resultados}
\end{figure*}

The behavior of the propagation velocity with respect to the thickness of the stabilizer is less obvious than the $v$ vs. $I$ behavior. Fig. \ref{resultados} shows a series of plots, where the propagation velocity is plotted as a function of the stabilizer thickness $L_2$ for different operating currents and magnetic fields. In each plot, the numerical results, the analytical approximation and the Wilson solution  are shown. Although no measurements on the behavior of the propagation velocity with respect to the geometry of the stabilizer are shown in the plots, we do expect, based on Fig. \ref{data}, that the results generally represent a correct behavior. 

Some insight can be gained by examining Fig. \ref{resultados}. For low currents, there is a very good agreement between our analytical and numerical models. For higher currents, this agreement breaks and some deviations between the two models appear. The general behavior of the plots can be explained, however, by examining the analytical formula, Eq. (\ref{vMB}). This general behavior is in fact similar to both models. The dynamics of the propagation velocity dependence on $L_2$ has three characteristics. 

\begin{enumerate}
\item First, for $L_2\ll1$ the dominant term is the current density, see Eq. (\ref{vMB}). As $L_2$ increases from zero, $J$ becomes smaller and leads to a small decrease in propagation velocity.

\item Second, when $\alpha < 1$, and for small enough values of $L_2$, the ongoing decrease in current density is compensated by a change in material properties, as the average values of the different material properties are more influenced by the presence of the Al stabilizer. Because of the large specific heat and thermal conductivity, and small resistivity, the cladding metal acts as a heat sink, therefore increasing the velocity.

\item Third, for large values of $L_2$, the cladding matrix becomes the dominant element in the conductor and practically determines the average material properties of the conductor. Hence, the material properties have practically a constant value and the Al stabilizer has fulfilled its potential for acting as a heat sink. Then, the current density, that goes down with $L_2$, becomes once more the dominant factor and thus reduces again the velocity.
\end{enumerate}

\section{Summary}
\label{conclusion}

The behavior of longitudinal normal zone propagation velocity was analyzed with respect to the thickness of the metal cladding of the Rutherford cable $L_2$ for a wide range of currents and magnetic fields. This provides a good estimation of the normal zone propagation for a variety of superconducting magnets. The results and the physics behind them were explained and analyzed. 

We have shown that for $\alpha < 1$ the current remains confined to a small area around the composite material, limiting the cladding's contribution to the normal zone propagation. Since the cladding is thick enough to act as a heat sink, the heat generated in the composite material quickly propagates into the cladding. This, in turn, accelerates the normal zone propagation because the heat generation that contributes to the normal zone propagation is formed almost exclusively within the Rutherford cable, which carries almost all of the current. 

Although we did not address the issue of minimum quench energy (MQE), we intend to do so in the future. Our results are not effected by this as the propagation velocity is constant for any initial energy release. In addition, our calculation can be expanded to a 3D and thus include the transverse velocities as well. 



\begin{thebibliography}{1}


\bibitem{ansysquench}
S. Caspi et al., Calculating quench propagation with ANSYS, \textit{IEEE Trans. Applied superconductivity} vol. 13, p. 1717 (2003). 

\bibitem{hermanquench}
H. H. J. ten Kate, H. Boschman and L. J. M. van de Klundert, Longitudinal propagation velocity of the normal zone in suerconducting wires, \textit{IEEE Trans. Magn.} vol. 23, p. 1557 (1987).

\bibitem{devred1}
A. Devred, General formulas for the adiabatic propagation velocity of the normal zone, \textit{IEEE Trans. Magn.} vol. 25, p. 1698 (1989).

\bibitem{willi}
M. Wilson, Superconducting Magnets, Oxford Press, New-York, USA (1983).

\bibitem{iaxojinst}
The IAXO collaboration, Conceptual Design of the International Axion Observatory (IAXO), JINST 9 T05002 (2014).

\bibitem{mints}
R. G. Mints, T. Ogitsu and A. Devred, Quench propagation velocity for highly stabilized conductors, Cryogenics, vol. 33, p. 449 (1992).

\bibitem{matpro}
L. Rossi and M. Sorbi, MATPRO: a computer library of material property at cryogenic temperature, CARE- Note-05-018-HHH (2005).

\bibitem{boxman}
E. W. Boxman, M. Pellegatta, A. V. Dudarev, and H. H. J. ten Kate, Current diffusion and normal zone propagation inside the aluminum stabilized superconductor of atlas model coil, \textit{IEEE Trans. Applied superconductivity}, vol. 13, p. 1685 (2003).



\bibitem{b00}
E. W. Boxman,  A. V. Dudarev, and H. H. J. ten Kate, The Normal Zone Propagation in ATLAS B00
Model Coil, \textit{IEEE Trans. Applied superconductivity}, vol. 12, p. 1549 (2002).

\bibitem{b0}
A. Foussat, N. Dolgetta, A. Dudarev, C. Mayri, P. Miele, Z. Sun, H. H. J. Ten Kate, and G. Volpini, Mechanical Characteristics of the ATLAS B0 Model Coil, \textit{IEEE Trans. Applied superconductivity}, vol. 13, p. 1246 (2003).

\end{thebibliography}
\end{document}